\documentclass{ceurart}
\usepackage{graphicx} 
\usepackage{subcaption}
\usepackage{bussproofs}
\usepackage{amsmath}
\usepackage{amsthm}
\usepackage{verbatim}
\usepackage{soul}
\usepackage{xcolor}
\usepackage{hyperref}
\usepackage{fancyvrb}
\usepackage{multirow}
\usepackage{makecell}
\usepackage{bussproofs}
\alwaysRootAtBottom
\definecolor{hlcolor}{HTML}{
    E3D7D4
}
\sethlcolor{hlcolor}
\usepackage{makecell}
\usepackage{todonotes}
\usepackage{makecell}
\usepackage{listings}
\usepackage[nomap, scaled=0.85]{FiraMono} 
\usepackage{tikz}
\usetikzlibrary{arrows, automata, shapes, positioning, tikzmark, patterns, calc, decorations, matrix}

\def\land{\wedge}

\def\lor{\vee}

\EnableBpAbbreviations

\newcommand{\goeland}{\textsf{Goéland}}
\newcommand{\coq}{\textsf{Coq}}
\newcommand{\dedukti}{\textsf{Dedukti}}
\newcommand{\lambdapi}{\textsf{Lambdapi}}
\newcommand{\gs}{\textsf{GS3}}
\newcommand{\zenon}{\textsf{Zenon}}
\newcommand{\zenonmodulo}{\textsf{ZenonModulo}}
\newcommand{\princess}{\textsf{Princess}}
\newcommand{\leantap}{\textsf{LeanTAP}}
\newcommand{\tptp}{\textsf{TPTP}}
\newcommand{\tstp}{\textsf{TSTP}}
\newcommand{\sctptp}{\textsf{SC-TPTP}}
\newcommand{\lisa}{\textsf{Lisa}}
\newcommand{\isabelle}{\textsf{Isabelle}}
\newcommand{\lean}{\textsf{Lean}}

  {\gdef\scalefactor{#1}\begin{center}\proofSkipAmount \leavevmode}%
  {\scalebox{\scalefactor}{\DisplayProof}\proofSkipAmount \end{center} }

\definecolor{dkgreen}{rgb}{0,0.6,0}
\definecolor{ltblue}{rgb}{0,0.4,0.4}
\definecolor{dkviolet}{rgb}{0.3,0,0.5}
\definecolor{green}{rgb}{0, 0.6, 0}

\lstdefinelanguage{scala}{
    alsoletter={@,=,>},
    keywordstyle = {\color{blue}},
    keywordstyle = [2]{\color{blue}},
    commentstyle = \color{comments},
    morekeywords = [2]{abstract, case, class, def, do, Input, Output, then,
        else, extends, false, free, if, implicit, match,
        object, true, val, var, while, sealed, or,
        for, dependent, null, type, with, try, catch, finally,
        import, final, return, new, override, this, trait,
        private, public, protected, package, throw},
    sensitive = true, 
    numbers=left,
    stepnumber=1,
    morecomment = [l]{//},
    morecomment = [s]{/*}{*/},
    morestring = [b]",  
    otherkeywords = {;,<<,>>,++},
    mathescape = true,
    escapeinside = {{*@}{@*}},
    literate = {
        {λ}{\(\lambda\)}{1}
    },
}

\lstdefinelanguage{lisa}{
    alsoletter={@,=,>},
    keywordstyle = {\color{blue}},
    keywordstyle = [2]{\color{blue}},
    keywordstyle = [3]{\color{green}},
    keywordstyle = [4]{\color{teal}},
    commentstyle = \color{comments},
    morekeywords = [2]{abstract, case, class, def, do, Input, Output, then,
        else, extends, false, free, if, implicit, match,
        object, true, val, var, while, sealed, or,
        for, dependent, null, type, with, try, catch, finally,
        import, final, return, new, override, this, trait,
        private, public, protected, package, throw},
    morekeywords = [3]{have, andThen, thenHave, Theorem, by, DEF, The, Lemma, subproof, assume},
    sensitive = true, 
    numbers=left,
    stepnumber=1,
    morecomment = [l]{//},
    morecomment = [s]{/*}{*/},
    morestring = [b]",  
    otherkeywords = {;,<<,>>,++},
}

\lstdefinelanguage{Coq}{ 
    mathescape=true,
    texcl=false, 
    morekeywords=[1]{Section, Module, End, Require, Import, Export,
        Variable, Variables, Parameter, Parameters, Axiom, Hypothesis,
        Hypotheses, Notation, Local, Tactic, Reserved, Scope, Open, Close,
        Bind, Delimit, Definition, Let, Ltac, Fixpoint, CoFixpoint, Add,
        Morphism, Relation, Implicit, Arguments, Unset, Contextual,
        Strict, Prenex, Implicits, Inductive, CoInductive, Record,
        Structure, Canonical, Coercion, Context, Class, Global, Instance,
        Program, Infix, Theorem, Lemma, Corollary, Proposition, Fact,
        Remark, Example, Proof, Goal, Save, Qed, Defined, Hint, Resolve,
        Rewrite, View, Search, Show, Print, Printing, All, Eval, Check,
        Projections, inside, outside, Def},
    morekeywords=[2]{forall, exists, exists2, fun, fix, cofix, struct,
        match, with, end, as, in, return, let, if, is, then, else, for, of,
        nosimpl, when},
    morekeywords=[3]{Type, Prop, Set, true, false, option},
    morekeywords=[4]{pose, set, move, case, elim, apply, clear, hnf,
        intro, intros, generalize, rename, pattern, after, destruct,
        induction, using, refine, inversion, injection, rewrite, congr,
        unlock, compute, ring, field, fourier, replace, fold, unfold,
        change, cutrewrite, simpl, have, suff, wlog, suffices, without,
        loss, nat_norm, assert, cut, trivial, revert, bool_congr, nat_congr,
        symmetry, transitivity, auto, split, left, right, autorewrite},
    morekeywords=[5]{by, done, exact, reflexivity, tauto, romega, omega,
        assumption, solve, contradiction, discriminate},
    morekeywords=[6]{do, last, first, try, idtac, repeat},
    morecomment=[s]{(*}{*)},
    showstringspaces=false,
    morestring=[b]",
    morestring=[d]’,
    tabsize=3,
    extendedchars=false,
    sensitive=true,
    breaklines=false,
    basicstyle=\small,
    captionpos=b,
    columns=[l]flexible,
    identifierstyle={\ttfamily\color{black}},
    keywordstyle=[1]{\ttfamily\color{violet}},
    keywordstyle=[2]{\ttfamily\color{green}},
    keywordstyle=[3]{\ttfamily\color{blue}},
    keywordstyle=[4]{\ttfamily\color{blue}},
    keywordstyle=[5]{\ttfamily\color{red}},
    stringstyle=\ttfamily,
    commentstyle={\ttfamily\color{dkgreen}},
    literate=
    {\\forall}{{\color{dkgreen}{$\forall\;$}}}1
    {\\exists}{{$\exists\;$}}1
    {<-}{{$\leftarrow\;$}}1
    {=>}{{$\Rightarrow\;$}}1
    {==}{{\code{==}\;}}1
    {==>}{{\code{==>}\;}}1
    {->}{{$\rightarrow\;$}}1
    {<->}{{$\leftrightarrow\;$}}1
    {<==}{{$\leq\;$}}1
    {\#}{{$^\star$}}1 
    {\\o}{{$\circ\;$}}1 
    {\@}{{$\cdot$}}1 
    {\/\\}{{$\wedge\;$}}1
    {\\\/}{{$\vee\;$}}1
    {++}{{\code{++}}}1
    {~}{{$\sim$}}1
    {\@\@}{{$@$}}1
    {\\mapsto}{{$\mapsto\;$}}1
    {\\hline}{{\rule{\linewidth}{0.5pt}}}1
}[keywords,comments,strings]

\conference{PAAR'24: 9th Workshop on Practical Aspects of Automated Reasoning,
      July 2, 2024, Nancy, France}
\title{SC-TPTP: An Extension of the TPTP Derivation Format for Sequent-Based Calculus}

\author[1]{Julie Cailler}[%
orcid=0000-0002-6665-8089,
email=julie.cailler@ur.de,
url=https://jcailler.github.io/,
]
\cormark[1]
\fnmark[1]
\address[1]{\textsf{University of Regensburg}, Regensburg, Germany}

\author[2]{Simon Guilloud}[%
orcid=0000-0001-8179-7549,
email=simon.guilloud@epfl.ch,
url=https://people.epfl.ch/simon.guilloud,
]
\cormark[1]
\fnmark[1]
\address[2]{\textsf{EPFL}, Lausanne, Switzerland}

\copyrightclause{Copyright for this paper by its authors. Use permitted under Creative Commons License Attribution 4.0 International (CC BY 4.0).}

\newtheorem{thm}{Theorem}[section]

\theoremstyle{definition}
\newtheorem{example}[thm]{Example}

\begin{document}

\begin{abstract}
    Motivated by the transfer of proofs between proof systems, and in particular from first order \textit{automated theorem provers} (ATPs) to \textit{interactive theorem provers} (ITPs), we specify an extension of the \tptp{} derivation~\cite{sutcliffeLogicLanguagesTPTP2023} text format to describe proofs in first-order logic: \sctptp{}. To avoid multiplication of standards, our proposed format over-specifies the \tptp{} derivation format by focusing on sequent formalisms. By doing so, it provides a high level of detail, is faithful to mathematical tradition, and cover multiple existing tools and in particular tableaux-based strategies. We make use of this format to allow the \lisa{} proof assistant~\cite{guilloudLISAModernProof2023} to query the Goéland automated theorem prover~\cite{caillerGoelandConcurrentTableauBased2022}, and implement a library of tools able to parse, print and check \sctptp{} proofs, export them into \coq{} files, and rebuild low-level proof steps from advanced ones.
\end{abstract}

\begin{keywords}
  TPTP \sep 
  Proof Format \sep
  Automated Theorem Proving \sep 
  Interactive Theorem Proving
\end{keywords}

\maketitle
\section{Introduction}

Transfer of proofs between different proof systems to this day largely remains a challenge. While the relative consistency strength of the foundations of most tools is well known, practical translations presents a variety of technical difficulties. Indeed, not only the syntax of different systems can be very hard (when not impossible) to simulate (for example, $\lambda$-abstractions used in type theory-based systems are difficult to represent and reason about in first-order systems, or an classical ATP and an intuitionistic ITP), but even proof systems with similar logical foundations can diverge significantly in terms of the kind and granularity of accepted \textit{proof steps}. In practice, an ATP might use advanced proof steps such as Skolemization, superposition, hyperresolution, or congruence closure, which can be difficult (in terms of implementation and space-time complexity) to simulate with lower-level proof steps typically accepted by proof assistants.
Finally, different tools may use different formats to export and store statements and proofs in the first place (e.g. \tptp~\cite{sutcliffeLogicLanguagesTPTP2023}, \textsf{XML}~\cite{kohlhaseExperiencesExportingMajor2021}, \textsf{LFSC}~\cite{barbosaCvc5VersatileIndustrialStrength2022}, \lambdapi~\cite{assaf2016dedukti} and other (tool-specific) formats), each requiring dedicated parsers and abstract syntax trees.

Hence, even when the foundations and proof steps are similar, if $n$ systems want to import proofs directly from each other, each is required to implement $n-1$ different import and proof transform algorithms, for a total of $n(n-1)$ implementations. With one common middle ground format, each system only needs to implement one import and one export algorithm from itself to the proof format, for a total of $2n$ implementations. On the other hand, as suggested above, unifying the syntax and proofs of arbitrarily many systems with unrelated foundations is overly ambitious and may not be practically feasible. A successful approach needs to convey the right level of abstraction, neither too specific nor too general.

\begin{figure}
    \centering
    \includegraphics[scale=0.5]{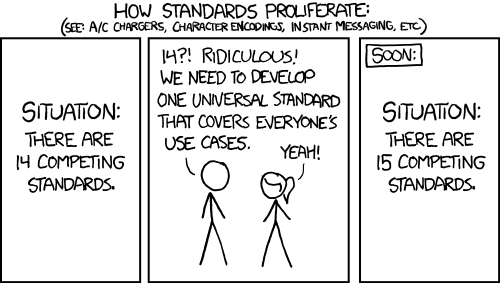}
    \caption{Standards (xkcd --- \url{https://xkcd.com/927/}). Licensed under CC BY-NC 2.5 License (\url{https://creativecommons.org/licenses/by-nc/2.5/})}
    \label{fig:xkcd}
\end{figure}

This introduction may remind the reader of the famous xkcd comics ``How Standards Proliferate'' (Figure~\ref{fig:xkcd}). Since being counterproductive runs against our objective, we design a proof format as a super specification of the \tptp{} derivation format, which is already supported by a large collection of systems and tools. \tptp{} (Thousands of Problems for Theorem Provers)~\cite{sutcliffeTPTPProblemLibrary2017} is a large library of problems for testing automated theorem provers. It is also a specification format~\cite{sutcliffeLogicLanguagesTPTP2023} for said problems and \textit{derivations}, i.e., lists of formulas derived from the specification of the problem or from previously deduced ones. However, there is no standard specification of \textit{how} formulas are derived.

Our motivation for the present proposal is to specifically increase interoperability between tools that use first-order logic. A key goal is to allow proof systems (in particular ITPs) to query ATPs on problems, and obtain proofs in a single format in return.
Hence, the present work defines a specification over the existing \tptp{} format for sequent-calculus style proofs in first-order logic. We fix and specify a set of basic inference steps, as well as their parameters, semantics, and syntax, in the style of sequent calculus. We separate deduction steps in \textit{levels}, from low-level, easy-to-verify standard steps of sequent calculus, to more complex and tool-specific steps, which ideally come with a procedure to transform them into low-level steps. We call the resulting standard \sctptp{}.  By keeping as-is \tptp's directives for everything else, including formulas, existing \tptp{} parsers and printers can be readily used. Our objectives are: 

\begin{itemize}
    \item to allow proof systems (in particular ITPs) using first-order logic to query ATPs for first-order logic (in particular tableaux-based ATPs) on problems, and obtain proofs in a single format in return;
    \item to provide a single translation exports from multiple systems with sequent calculus-like foundations to another proof system or format (for example \coq);
    \item to offer transformation algorithms that simplify proofs with possibly complicated and higher level proof steps to proofs with only basic steps;
    \item to allow answers of ATPs in competitions to be verified unambiguously (for problems in first-order logic).
\end{itemize}

\paragraph{Contributions}
Our contributions in the present text are as follows:
\begin{itemize}
    \item The development of a standard format, \sctptp{}, for sequent-based calculus, with a focus on tableaux-based ATP.
    \item The implementation of this format into the \lisa{} proof assistant (import), and the \goeland{} ATP (export). In particular, this allows \lisa{} users to call \goeland{} as a (proof-producing) tactic.
    \item The development of a publicly available library, providing tools to parse and print \sctptp{} proofs, check their validity in this format, transform proofs to eliminate high-level congruence closure steps, and export them into \coq. 
\end{itemize}

This work is a proof of concept and a proposal to the community, that we hope can help connect various tools. We look forward to the community's input and feedback on how to adapt the features and design choices to be suitable for as many tools as possible.

\paragraph{Related Work}

Beyond the \tptp{} format itself \cite{sutcliffeLogicLanguagesTPTP2023} and its associated body of work, significant research efforts have been directed toward designing proof formats conducive to tool interoperability. Those efforts started by conducting various investigations in order to delineate the criteria for an ideal proof format, emphasizing the importance of ease of parsing while retaining human readability \cite{bohme2011designing, reger2017checkable}.

One of the original inspirations for designing a common format comes from the SAT community, which encountered notable success with the \textsf{DRAT} format~\cite{wetzler2014drat}. Work has also been done to provide a proof checker for this format~\cite{cruz2017efficient}.

In the realm of SMT solving, diverse proof formats have emerged, including \textsf{LFSC}~\cite{stump2008towards} used by \textsf{CVC5}~\cite{barbosa2022cvc5} or the \textsf{Z3}~\cite{de2008z3} proof format~\cite{de2008proofs}, among others. While there is currently no universal proof format for SMT solvers, efforts are made toward this goal~\cite{besson2011flexible, stump2008towards, hoenicke2022simple}, resulting for instance in the \textsf{Alethe}~\cite{ schurr2021alethe} proof format, employed by the \textsf{VeriT} solver \cite{bouton2009verit}.

Closer to our approach, and extension of the \tptp{} syntax to the connection calculus~\cite{otten2023syntax} was implemented into the \textsf{leanCoP}~\cite{otten2003leancop} and \textsf{Connect++}~\cite{holdenConnectNewAutomated2023} provers. In the meantime, the Theory Extensible Sequent Calculus (\textsf{TESC}) format for First-Order ATPs \cite{baek2020tesc} offers a sequent-based proof format capable of compiling and verifying solutions from \textsf{Vampire}~\cite{kovacs2013first} and \textsf{E}~\cite{schulz2019faster} by combining \tptp{} problems with their respective \tstp{} solutions. Within the \tptp{} ecosystem, the  \textsf{GDV} Verifier~\cite{sutcliffe2006semantic} proof checker for CNF derivation in the \tptp{} format warrant mention.

Finally, our paper's overarching theme resides in the realm of proof interoperability, echoing the ethos of systems such as \dedukti{} and \lambdapi~\cite{assaf2016dedukti}.

\paragraph{Organization}
In \autoref{sect:context}, we provide the necessary background on \tptp, tableaux rules, as well as one-sided and two-sided sequent calculi. We then introduce the proof steps of the \sctptp{} proof standard for (untyped) first order-logic in \autoref{sect:technicalDescription}, together with their parameters, syntax, and semantic. In \autoref{sect:utilities}, we present a library of utilities to handle and verify proofs in \sctptp{}. In particular, we implement an export of \sctptp{} proofs to \coq{}, and a proof-producing egraph~\cite{willseyEggFastExtensible2021} to produce detailed proofs corresponding to congruence closure steps.
Finally, we describe \autoref{sect:interoperability_experiments}, as a use case and proof of concept, how we made the \goeland{} ATP and the \lisa{} ITP support \sctptp{}, allowing for direct transfer of proofs from the first to the second.

\section{Context}
\label{sect:context}

\subsection{\tptp{} \& \tstp{}}

\tptp{}~\cite{sutcliffeTPTPProblemLibrary2017} is the reference problem library in the field of automated reasoning. It is made of over 25000 problems (including over 7000 first-order (\texttt{FOF} category) problems), ranging from easy to open, that can be used for testing and evaluating ATPs. This library also provides standards for input and output for ATP systems, thanks to the \tptp{} logic language~\cite{sutcliffeLogicLanguagesTPTP2023}, that is used to specify logical decision problems in various logic ((typed) first-order, higher order, each with and without polymorphism, CNFs, ...). It is a well-adopted input format and benchmark, used for example in CASC, the CADE ATP System Competition. 


Sibling to \tptp{} is the \tstp{} library, a collection of solutions (derivations, or proofs) produced by ATPs in response to \tptp{} problems. In derivations, formulas are annotated by an \textit{inference} parameter, itself made of the name of the inference rule, its parameter, and its premises. However, inferences are not specified: every system can output its individual proof steps, which can be arbitrarily complex to reconstruct for a system with different or more basic deduction rules. As an extreme example, an ATP may output a proof step justified with ``SMT solver said so'', from which it would be very difficult to recover a posteriori a syntactically strict proof, such as a sequent calculus or natural deduction proof or a well-typed proof term. Even when systems output reasonable, low-level steps, there can be many inessential differences in the exact set of rules and their syntax.

\subsection{Sequent-Based Calculus}
Sequent Calculus is a proof system for (classical) first-order logic (with equality), where statements are represented by sequents. A sequent is a pair of sets of formulas, typically written
$$
a_1,...,a_n \vdash b_1,...,b_n
$$
whose intended semantics is $(a_1 \land ... \land a_n ) \implies (b_1\lor ...\lor b_n)$. 

\paragraph{LK \& LJ} The original sequent calculus LK~\cite{gentzenUntersuchungenUberLogische1935} for classical logic, and its counterpart LJ~\cite{gentzen1935untersuchungen} for intuitionistic logic, are formals system designed to study natural deduction in first-order logic. Sequent calculus admits a \textit{subformula property}, which gives it its high theoretical and practical relevance. In addition, this calculus also provides a naive (but very inefficient in practice) complete proof search procedure, on which are built more refined proof-search strategies such as the method of analytics tableaux.

\paragraph{Tableaux}
The tableaux method is a semi-decision procedure for first-order logic. A variety of ATPs are based on the (free-variable) tableaux method, such as \princess~\cite{rummerConstraintSequentCalculus2008}, \goeland~\cite{caillerGoelandConcurrentTableauBased2022}, \zenon~\cite{bonichonZenonExtensibleAutomated2007}, \zenonmodulo~\cite{delahayeZenonModuloWhen2013}, \leantap~\cite{beckertLeanTAPLeanTableaubased1994} and others. 
One of the advantages of this method is that it follows generally the proof steps of sequent calculus, making it suitable for the reconstruction of efficiently and independently verifiable proofs. The tableaux calculus consists of a set of (refutationally complete) inference rules, divided in four categories: $\alpha$ rules (unary inferences), $\beta$ rules (binary inferences), $\gamma$ rules (instantiation rules) and $\delta$ rules (Skolemization rules).

\paragraph{GS3}
The Gentzen-Schütte calculus (\gs{})~\cite{troelstraBasicProofTheory2000} is a one-sided variant of sequent calculus where formulas are only allowed on the left side. The deduction rules are similar to those of the usual sequent calculus, but the right rules are replaced by left-negation rules. Except for equality, it is easy to observe that the three proof systems are equivalent, with GS3 as a middle ground. For equality reasoning, however, expressing the global closure substitution step using the equality substitution rules is non-trivial, and will be the topic of \autoref{sect:utilities}.

\section{\sctptp{}: Technical Description}
\label{sect:technicalDescription}

We introduce the \sctptp{} format as an over-specification of the \tptp{} format~\cite{sutcliffeLogicLanguagesTPTP2023} for derivations, with two level of derivation steps. A derivation is a list of \textit{annotated formulas}, which can be conjectures or axioms as in the specification of \tptp{} problems, or derived from previously annotated formulas via some inference rule and parameters. Formulas themselves are part of the FOFX grammar of \tptp{} (Figure~\ref{fig:fofx}), which extends FOF by supporting notation for sequents. Until now, the FOFX format was defined but ``not yet in use''. Technically, TPTP does not allow free variables. However, they are an integral part of sequent calculus\footnote{and in general of deductive reasoning, for example in the proof system of HOL Light~\cite{harrisonHOLLightOverview2009} and in informal mathematics}, and the grammar supports free variables.

 \begin{figure}[tbhp]
        \parbox{\textwidth}{
            \begin{align*}
                \texttt{<fof\_annotated>} ::= & \qquad \texttt{fof(<name>, <formula\_role>, <fof\_formula>, <annotations>).}            \\
                \texttt{<formula\_role>} ::= & \qquad \texttt{assumption | axiom | conjecture | plain}  \\ 
                \texttt{<fof\_formula>} ::= & \qquad \texttt{<fof\_logical\_formula>}  \mid \texttt{<fof\_sequent>}  \\ 
                \texttt{<fof\_sequent>} ::= & \qquad \texttt{<fof\_formula\_tuple>  \texttt{-->}  <fof\_formula\_tuple>}                        \\
                \mid     \;         & \qquad \texttt{(<fof\_sequent>)}  \\ 
                \texttt{<fof\_formula\_tuple>} ::= & \qquad \texttt{[]} \;  \mid  \texttt{ [<fof\_formula\_tuple\_list>] }                      \\
                \texttt{<fof\_formula\_tuple\_list>} ::= & \qquad \texttt{ <fof\_logic\_formula> } \\
                 \mid     \;         & \qquad \texttt{<fof\_logic\_formula>, <fof\_formula\_tuple\_list>}  
            \end{align*}}
    \caption{Main elements of the \sctptp{} syntax. \texttt{logic\_formula} is the FOF syntax for logical formulas. For \sctptp{} derivations, annotations are of the form \texttt{inference(rule, [parameters], [premises])}.}
    \label{fig:fofx}
\end{figure}

In the original presentation of Gentzen, sequents' sides are formally \textit{lists} of formulas, where order and number of duplicates are significant. This semantics requires additional structural rules for contraction and permutation of formulas. However, it is more convenient and efficient (in terms of proof size and number of rules) to consider them as sets. This is the semantic we chose for \sctptp{}. In particular, formulas in sequents can be duplicated, contracted, and reordered at will. This does not make proof checking more complex.

\renewcommand{\arraystretch}{1.80} 

\begin{table}[th]
\small
    \centering
    \begin{tabular}{|c|c|c|l|}
        \hline 
        Rule name & Premises & Rule & Parameters \\ \hline 
        \texttt{hyp}       &  0 &  \makecell{\AxiomC{} \UnaryInfC{$\Gamma, A \vdash A, \Delta$} \DisplayProof}  & \makecell[l]{\texttt{i:Int}: Index of $A$ on the left  \\ \texttt{j:Int}: Index of $A$ on the right} \\ \hline 
        \texttt{leftHyp}  & 0 &  \makecell{\AxiomC{} \UnaryInfC{$\Gamma, A, \neg A \vdash \Delta$} \DisplayProof}  & \makecell[l]{\texttt{i:Int}: Index of $A$ on the left  \\ \texttt{j:Int}: Index of $\neg A$ on the left.} \\ \hline
        \texttt{leftWeaken}  & 1 &\makecell{\AxiomC{$\Gamma \vdash \Delta$} \UnaryInfC{$\Gamma, A \vdash \Delta$} \DisplayProof}          & \texttt{i:Int}: Index of $A$ on the left  \\ \hline 
        \texttt{rightWeaken} & 1 &\makecell{\AxiomC{$\Gamma \vdash \Delta$} \UnaryInfC{$\Gamma \vdash A, \Delta$} \DisplayProof}      & \texttt{i:Int}: Index of $A$ on the right \\ \hline 
        \texttt{cut}         & 2 & \makecell{\AxiomC{$\Gamma \vdash A, \Delta$}\AxiomC{$\Sigma, A \vdash \Pi$}\BinaryInfC{$\Gamma, \Sigma \vdash \Delta, \Pi$} \DisplayProof} & \makecell[l]{\texttt{i:Int}: Index of the cut formula on the\\right of the first premise} \\ \hline 
    \end{tabular}
    \caption{Level 1 rules of \sctptp, for one-sided and two-sided sequent calculus --- structural rules.}
    \label{tab:SCTPTP_rules_lvl_1_1}
\end{table}

\begin{table}[th]
\small
    \centering
    \begin{tabular}{|c|c|c|l|}
         \hline
        Rule name & Premises & Rule & Parameters \\ \hline
        \texttt{leftAnd}     & 1 & \makecell{\AxiomC{$\Gamma, A, B \vdash \Delta$} \UnaryInfC{$\Gamma, A \land B \vdash \Delta$} \DisplayProof} & \texttt{i:Int}: Index of $A \land B$ on the left    \\ \hline
        \texttt{leftOr}      & 2 & \makecell{\AxiomC{$\Gamma, A \vdash \Delta$} \AxiomC{$\Sigma, B \vdash \Pi$} \BinaryInfC{$\Gamma, \Sigma, A \lor B \vdash \Delta, \Pi$} \DisplayProof}    & \texttt{i:Int}: Index of $A \lor B$ on the left     \\ \hline
        \texttt{leftImp1}    & 2 & \makecell{\AxiomC{$\Gamma \vdash A, \Delta$}\AxiomC{$\Sigma, B \vdash \Pi$}\BinaryInfC{$\Gamma, \Sigma, A \Rightarrow B \vdash \Delta, \Pi$} \DisplayProof} & \texttt{i:Int}: Index of $A \Rightarrow B$ on the left \\ \hline
        \texttt{leftImp2}    & 2 & \makecell{\AxiomC{$\Gamma, \neg A \vdash \Delta$}\AxiomC{$\Sigma, B \vdash \Pi$}\BinaryInfC{$\Gamma, \Sigma, A \Rightarrow B \vdash \Delta, \Pi$} \DisplayProof} & \texttt{i:Int}: Index of $A \Rightarrow B$ on the left \\ \hline
        \texttt{leftIff}     & 1 & \makecell{\AxiomC{$\Gamma, A \Rightarrow B, B \Rightarrow A \vdash \Delta$}\UnaryInfC{$\Gamma, A \Leftrightarrow B \vdash \Delta$}\DisplayProof} & \texttt{i:Int}: Index of $A \Leftrightarrow B$ on the left     \\ \hline
        \texttt{leftNot}     & 1 & \makecell{\AxiomC{$\Gamma \vdash A, \Delta$}\UnaryInfC{$\Gamma, \neg A \vdash \Delta$}\DisplayProof } & \texttt{i:Int}: Index of $\neg A$ on the left       \\ \hline
        \texttt{leftEx}      & 1 & \makecell{\AxiomC{$\Gamma, A \vdash \Delta$}\UnaryInfC{$\Gamma, \exists x. A \vdash \Delta$}\DisplayProof} & \makecell[l]{\texttt{i:Int}: Index of $\exists x. A$ on the left      \\ \texttt{y:Var}: Variable in place of $x$ in the premise} \\ \hline
        \texttt{leftAll}     & 1 & \makecell{\AxiomC{$\Gamma, A[x := t] \vdash \Delta$}\UnaryInfC{$\Gamma, \forall x. A  \vdash \Delta$} \DisplayProof} & \makecell[l]{\texttt{i:Int}: Index of $\forall x. A$ on the left      \\ \texttt{t:Term}: Term in place of $x$ in the premise} \\ \hline
    \end{tabular}
    \caption{Level 1 rules of \sctptp, for one-sided and two-sided sequent calculus --- left introduction rules.}
    \label{tab:SCTPTP_rules_lvl_1_2}
\end{table}

\begin{table}[th]
\small
    \centering
    \begin{tabular}{|c|c|c|l|}
         \hline
        Rule name & Premises & Rule & Parameters \\ \hline
        \texttt{rightAnd}     & 2 &  \makecell{\AxiomC{$\Gamma \vdash A, \Delta$} \AxiomC{$\Sigma \vdash B, \Pi$} \BinaryInfC{$\Gamma, \Sigma \vdash  A \land B, \Delta, \Pi$} \DisplayProof} & \texttt{i:Int}: Index of $A \land B$ on the right    \\ \hline
        \texttt{rightOr}      & 1 &  \makecell{\AxiomC{$\Gamma \vdash A, B, \Delta$} \UnaryInfC{$\Gamma \vdash A \lor B, \Delta$} \DisplayProof}  & \texttt{i:Int}: Index of $A \lor B$ on the right     \\ \hline
        \texttt{rightImp}    & 1 & \makecell{\AxiomC{$\Gamma, A \vdash B, \Delta$}\UnaryInfC{$\Gamma \vdash A \Rightarrow B, \Delta$} \DisplayProof} & \texttt{i:Int}: Index of $A \Rightarrow B$ on the right \\ \hline
        \texttt{rightIff}    & 2 & \makecell{\AxiomC{$\Gamma \vdash A \Rightarrow B, \Delta$}\AxiomC{$\Sigma \vdash B \Rightarrow A, \Pi$}\BinaryInfC{$\Gamma, \Sigma \vdash A \Leftrightarrow B, \Delta, \Pi$} \DisplayProof} & \texttt{i:Int}: Index of $A \Leftrightarrow B$ on the right \\ \hline
        \texttt{rightNot}     & 1 & \makecell{\AxiomC{$\Gamma, A \vdash \Delta$}\UnaryInfC{$\Gamma \vdash \neg A, \Delta$}\DisplayProof } & \texttt{i:Int}: Index of $\neg A$ on the right       \\ \hline
        
        \texttt{rightEx}     & 1 & \makecell{\AxiomC{$\Gamma, A[x := t] \vdash \Delta$}\UnaryInfC{$\Gamma, \exists x. A  \vdash \Delta$} \DisplayProof} & \makecell[l]{\texttt{i:Int}: Index of $\exists x. A$ on the right      \\ \texttt{t:Term}: Term in in place of $x$ in the\\premise} \\ \hline
        \texttt{rightAll}      & 1 & \makecell{\AxiomC{$\Gamma, A[x := y] \vdash \Delta$}\UnaryInfC{$\Gamma, \forall x. A \vdash \Delta$}\DisplayProof} & \makecell[l]{\texttt{i:Int}: Index of $\forall x. A$ on the right      \\ \texttt{y:Var}: Variable in place of $x$ in the\\premise} \\ \hline
    \end{tabular}
    \caption{Level 1 rules of \sctptp, for one-sided and two-sided sequent calculus --- right introduction rules.}
    \label{tab:SCTPTP_rules_lvl_1_3}
\end{table}

\begin{table}[bh]
\small
\centering
 \begin{tabular}{|c|c|c|l|}
         \hline
        Rule name & Premises & Rule & Parameters \\ \hline
        \texttt{leftNotAnd}  & 2 & \makecell{\AxiomC{$\Gamma, \neg A \vdash \Delta$} \AxiomC{$\Sigma, \neg B \vdash \Pi$} \BinaryInfC{$\Gamma, \Sigma, \neg(A \land B) \vdash \Delta, \Pi$} \DisplayProof}     & \makecell[l]{\texttt{i:Int}: Index of $\neg(A \land B)$\\ on the left}     \\ \hline
        \texttt{leftNotOr}   & 1 & \makecell{\AxiomC{$\Gamma, \neg A, \neg B \vdash \Delta$} \UnaryInfC{$\Gamma, \neg(A \lor B) \vdash \Delta$} \DisplayProof} & \makecell[l]{\texttt{i:Int}: Index of $\neg(A \lor B)$\\ on the left}      \\ \hline
        \texttt{leftNotImp}  & 1 & \makecell{\AxiomC{$\Gamma, A, \neg B \vdash \Delta$} \UnaryInfC{$\Gamma, \neg(A \Rightarrow B) \vdash \Delta$} \DisplayProof} & \makecell[l]{\texttt{i:Int}: Index of $\neg(A \Rightarrow B)$\\on the left}  \\ \hline
        \texttt{leftNotIff}  & 2 & \makecell{\AxiomC{$\Gamma, \neg (A \Rightarrow B) \vdash \Delta$} \AxiomC{$\Sigma, \neg (B \Rightarrow A) \vdash \Pi$} \BinaryInfC{$\Gamma, \Sigma, \neg(A \Leftrightarrow B) \vdash \Delta, \Pi$} \DisplayProof} & \makecell[l]{\texttt{i:Int}: Index of $\neg(A \Leftrightarrow B)$\\on the left}      \\ \hline
        \texttt{leftNotNot}  & 1 & \makecell{\AxiomC{$\Gamma, A \vdash \Delta$} \UnaryInfC{$\Gamma, \neg \neg A \vdash \Delta$} \DisplayProof} & \makecell[l]{\texttt{i:Int}: Index of $\neg \neg A$\\on the left}         \\ \hline
        \texttt{leftNotEx}   & 1 & \makecell{\AxiomC{$\Gamma, \neg A[x := t] \vdash \Delta$}\UnaryInfC{$\Gamma, \neg\exists x. A  \vdash \Delta$} \DisplayProof}  & \makecell[l]{\texttt{i:Int}: Index of $\neg \exists x. A$\\ on the right     \\  \texttt{t:Term}: Term in place of $x$\\ in the premise}   \\ \hline
        \texttt{leftNotAll}  & 1 & \makecell{\AxiomC{$\Gamma, \neg A \vdash \Delta$}\UnaryInfC{$\Gamma, \neg \forall x. A \vdash \Delta$}\DisplayProof} & \makecell[l]{\texttt{i:Int}: Index of $\neg \forall x. A$\\ on the right\\ \texttt{y:Var}: Variable in place of\\$x$ in the premise} \\ \hline
    
    \end{tabular}
    \caption{Level 1 rules of \sctptp, for one-sided and two-sided sequent calculus --- left not introduction rules. These rules are equivalent to the right rules of \autoref{tab:SCTPTP_rules_lvl_1_3}, but more directly match tableaux reasonning.}
    \label{tab:SCTPTP_rules_lvl_1_4}
\end{table}

\begin{table}[bh]
\small
\centering
 \begin{tabular}{|c|c|c|l|}
         \hline
        Rule name & Premises & Rule & Parameters \\ \hline 
        \texttt{rightRefl}  & 0 & \makecell{\AxiomC{} \UnaryInfC{$\Gamma \vdash t = t, \Delta$} \DisplayProof}  & \texttt{i:Int}: Index of $ t = t $ on the right. \\ \hline
        \texttt{rightSubst}  & 1 &  \makecell{\AxiomC{$\Gamma, t = u \vdash P(t), \Delta$} \UnaryInfC{$\Gamma, t = u \vdash P(u), \Delta$} \DisplayProof}  & \makecell[l]{\texttt{i:Int}: Index of $t = u$ on the left \\ \texttt{P(Z):Var}: Shape of the predicate on the right \\ \texttt{Z:Var}: unifiable sub-term in the predicate} \\ \hline                   
        \texttt{leftSubst}  & 1 & \makecell{\AxiomC{$\Gamma, t = u, P(t) \vdash \Delta$} \UnaryInfC{$\Gamma, t = u,  P(u) \vdash \Delta$} \DisplayProof} & \makecell[l]{\texttt{i:Int}: Index of $t = u$ on the left   \\ \texttt{P(Z):Term}: Shape of the predicate on the left \\ 
        \texttt{Z:Var}: variable indicating where to substitute} \\\hline    
    \end{tabular}
    \caption{Level 1 rules of \sctptp, for one-sided and two-sided sequent calculus --- equality reasoning.}
    \label{tab:SCTPTP_rules_lvl_1_5}
\end{table}

\renewcommand{\arraystretch}{1}

\subsection{Derivations}
Each step of the derivation is written as an \emph{annotated statement} of the form \texttt{fof(\textit{name},\textit{role},\textit{statement},\textit{annotation})}, in which: 
\begin{itemize}
    \item \texttt{\textit{name}} is an integer or an alphanumeric identifier starting with a lowercase letter. It is used to be referred to by other steps of the derivation.
    \item \texttt{\textit{role}} is either ``axiom'', ``conjecture'', ``assumption'' or ``plain''. An ``axiom'' denotes an accepted formula, while a ``conjecture'' holds for the statement the derivation is supposed to prove, but does not play a logical role. A conjecture should always contain the same formula as the last derived formula in the proof. The \tptp{} syntax does not impose specific user semantics to the ``plain'' role, and thus we use it to denote inferred steps.
    \item \texttt{\textit{statement}} can be either of a sequent or a formula. Their syntax can be found in \url{https://tptp.org/TPTP/SyntaxBNF.html}. For \sctptp{}, a \emph{sequent} statement is made of two sets of formulas, while a \emph{formula} statement is understood as standing for the sequent with an empty left-hand side and whose right-hand side contains exactly this formula.
\item \texttt{\textit{annotation}} are used to give additional information to the system. If the role of the statement is ``axiom'' or ``conjecture'', the annotation has no specific requirement in our format. For ``assumption'' and ``plain'' statements, the \texttt{\textit{annotation}} must be of the form \lstinline[mathescape]|inference(stepName, [status(thm), p${}_1$, ..., p${}_n$], [r${}_1$, ..., r${}_n$])|, in which:
    \begin{itemize}
        \item \texttt{\textit{stepName}} is one of the entry listed in \autoref{tab:SCTPTP_rules_lvl_1_1}-\autoref{tab:SCTPTP_rules_lvl_2}.
        \item \lstinline|status(thm)| indicates the status of the formula (e.g., a consequence of the premise, equisatisfiable with regard to the previous step, a negated conjecture, etc.) following the \textit{SZS ontologies}~\cite{sutcliffe2008szs}. In SC-TPTP, all steps are deductive inference and hence we only use the \lstinline|thm| status.
        \item The elements \texttt{p}${}_i$'s within the parameter list vary in number and shape based on the proof step (indexes, variables, (first-order) terms (\texttt{\$fot}), etc). They are described in the  4\textsuperscript{th} column of \autoref{tab:SCTPTP_rules_lvl_1_1}-\autoref{tab:SCTPTP_rules_lvl_2} and typically indicate how the step is constructed, thus making its correctness easily verifiable, without requiring inference. 
        \item The elements \texttt{r}${}_i$'s in the premises list point to the premises of the deduction step. Their number varies between 0 and 2, depending on the proof step, and are referenced in the 2\textsuperscript{nd} column of \autoref{tab:SCTPTP_rules_lvl_1_1}-\autoref{tab:SCTPTP_rules_lvl_2}.
    \end{itemize}
\end{itemize}

\begin{example}
    The following example illustrates a valid \sctptp{} derivation. 
    \begin{lstlisting}
fof(c, conjecture, [![X]: P(X)] --> [P(A) & P(B)]).
fof(s1, assumption, [P(A)] --> [P(A)], inference(hyp, [status(thm), 0, 0], [])).
fof(s2, plain, [![X]: P(X)] --> [P(A)], 
inference(leftAll, [status(thm), 0, $fot(A)], [s1])).
fof(s3, assumption, [P(B)] --> [P(B)], inference(hyp, [status(thm), 0, 0], [])).
fof(s4,plain, [![X]: P(X)] --> [P(B)], 
inference(leftAll, [status(thm), 0, $fot(B)], [s3])).
fof(s5,plain, [![X]: P(X)] --> [P(A) & P(B)], 
inference(rightAnd, [status(thm), 0], [s2, s4])).
    \end{lstlisting}
\end{example}

\begin{example}
    \sctptp{} proof of the \emph{drinker paradox}~\cite{smullyan1978name}.
\begin{lstlisting}
fof(c_drinkers_p, conjecture, (? [X] : d(X) => (! [Y] : d(Y)))).

fof(f8, assumption, [~(? [X] : d(X) => (! [Y] : d(Y))), 
~(d(X_0) => (! [Y] : d(Y))), d(X_0), ~(! [Y] : d(Y)), ~d(Sko_0), 
~(d(Sko_0) => (! [Y] : d(Y))), d(Sko_0)] --> [], 
inference(leftHyp, [status(thm), 6, 4], [])).

fof(f7, plain, [~(? [X] : d(X) => (! [Y] : d(Y))), 
~(d(X_0) => (! [Y] : d(Y))), d(X_0), ~(! [Y] : d(Y)), ~d(Sko_0), 
~(d(Sko_0) => (! [Y] : d(Y)))] --> [], inference(leftNotImp, [status(thm), 5], [f8])).

fof(f6, plain, [~(? [X] : d(X) => (! [Y] : d(Y))), 
~(d(X_0) => (! [Y] : d(Y))),  d(X_0), ~(! [Y] : d(Y)), ~d(Sko_0)] 
--> [], inference(leftNotEx, [status(thm), 0, $fot(Sko_0)], [f7])).

fof(f5, plain, [ ~(? [X] : d(X) => (! [Y] : d(Y))), 
~(d(X_0) => (! [Y] : d(Y))),  d(X_0),  ~(! [Y] : d(Y))] --> [], 
inference(leftNotForall, [status(thm), 3, $fot(Sko_0)], [f6])).

fof(f4, plain, [~(? [X] : d(X) => (! [Y] : d(Y))), 
~(d(X_0) => (! [Y] : d(Y)))] --> [], inference(leftNotImp, [status(thm), 1], [f5])).

fof(f3, plain, [~(? [X] : d(X) => (! [Y] : d(Y)))] --> [], 
inference(leftNotEx, [status(thm), 0, $fot(X_0)], [f4])).

fof(f2, assumption, [(? [X] : d(X) => (! [Y] : d(Y)))] --> 
[(? [X] : d(X) => (! [Y] : d(Y)))], inference(hyp, [status(thm), 0, 0], [])).

fof(f1, plain, [] --> [(? [X] : d(X) => (! [Y] : d(Y))), 
~(? [X] : d(X) => (! [Y] : d(Y)))], inference(rightNot, [status(thm), 1], [f2])).

fof(f0, plain, [] --> [(? [X] : d(X) => (! [Y] : d(Y)))], 
inference(cut, [status(thm), 0], [f1, f3])).
    \end{lstlisting} 
\end{example}

\subsection{Level 1 Deduction Steps}
We define three \textit{levels} for deduction steps. Level 1 steps are exactly the 30 steps represented in \autoref{tab:SCTPTP_rules_lvl_1_1} to \autoref{tab:SCTPTP_rules_lvl_1_5}. In this setup, ``on the left/right'' refers to the left and the right of the conclusion. Those rules are complete for first order logic with equality, and their correctness is simple to check. As they represents low-level proof steps they should also be straightforward to import into any proof system strong enough to accommodate first order logic. They encompass both the traditional two-sided sequent calculus rules, with antecedents and succedants, as well as the rules for one-sided sequent calculus typically used for tableaux theorem proving. As such, the system is not minimal.

\autoref{tab:SCTPTP_rules_lvl_1_1} presents the first group of rules (\texttt{hyp}, \texttt{leftHyp}, \texttt{leftWeaken}, \texttt{rightWeaken}, \texttt{cut}), also called \textit{structural rules}. The next group of 8 rules introduced in \autoref{tab:SCTPTP_rules_lvl_1_2} is composed by \textit{left introduction rules}, all of them describing a specific way a symbol can be introduced. The subsequent group in \autoref{tab:SCTPTP_rules_lvl_1_3} encompass \textit{right introduction rules}, dual to the left rules. Then follow \textit{left not introduction rules} in \autoref{tab:SCTPTP_rules_lvl_1_4}, which are essentially equivalent to the right introduction rules, but useful for tableaux-style proofs. Finally, \texttt{rightRefl}, \texttt{leftSubst} and \texttt{rightSubst} of \autoref{tab:SCTPTP_rules_lvl_1_5}  support equality reasoning.

\subsubsection*{Parameters and steps correctness}
In an \sctptp{} derivation, inferences with level 1 steps come with a set of parameters that makes verification straightforward and more efficient, without requiring inference. For most proof steps, the parameters are only indexes, pointing to the position of the formula targeted by the proof step. For example, consider a derivation with the \texttt{rightAnd} rule:
\begin{lstlisting}
fof(ax1, axiom, [] --> [a])
fof(ax2, axiom, [] --> [b])
fof(s1, plain, [] --> [a & b], inference(rightAnd, [status(thm), 0], [ax1, ax2]))
\end{lstlisting}
The step \lstinline|s1| is inferred using the rule \texttt{rightAnd}. The index $0$ in \lstinline|[0]| indicates that the deduced conjunction is the first formula (on the right-hand side of the conclusion sequent), that is \lstinline|a & b|. This then indicates that \lstinline|a| is a formula on the right of the first premise (\lstinline|ax1|) and \lstinline|b| a formula on the right of the second premise (\lstinline|ax2|). In the general case, let $\Gamma_1 \vdash \Delta_1$, $\Gamma_2 \vdash \Delta_2$ and $\Gamma_3 \vdash \Delta_3$ be the sequents of \lstinline|ax1|, \lstinline|ax2| and \lstinline|s1|. To verify that the step is correctly applied, we simply have to check the following conditions:
$$\Gamma_3 == \Gamma_1 \cup \Gamma_2$$
$$\lbrace \text{\lstinline|a|}\rbrace \cup \Delta_3 == \lbrace \text{\lstinline|a & b|}\rbrace \cup \Delta_1$$
$$\lbrace \text{\lstinline|b|}\rbrace \cup \Delta_3 == \lbrace \text{\lstinline|a & b|}\rbrace \cup \Delta_2$$
Where $==$ is equality on sets. Note that the $\Delta$'s may contain additional occurrences of \lstinline|a|, \lstinline|b| and \lstinline|a & b|, in which case the step is still valid.
All \lstinline|left|, \lstinline|right| and \lstinline|leftNot| propositional steps, as well as \lstinline|rightRefl| and all structural steps but \lstinline|cut| work the same way. Note that using hash sets, these tests can be done in time linear in the size of the sequents.

\lstinline|Ex| and \lstinline|All| steps take an additional argument, denoting which subterm or variable is being quantified. Consider:
\begin{lstlisting}
fof(s1,plain, [] --> [(f(X) = f(X))], 
        inference(rightRefl, [status(thm), 0], []))
        
fof(s2,plain, [] --> [?[Y]: (f(X) = Y)], 
        inference(rightEx, [status(thm), 0, $fot(f(X))], [s1]))
        
fof(s3,plain, [] --> [![X]: (?[Y]: (f(X) = Y))], 
        inference(rightAll, [status(thm), 0, $fot(X)], [s1]))
\end{lstlisting}
For \lstinline|s2|, the parameter $0$ indicates that the first formula in the right of the conclusion has been quantified, which is \lstinline|?[Y]: (f(X) = Y)|. Again in a more general case with arbitrary contexts, let $\Gamma_1 \vdash \Delta_1$, $\Gamma_2 \vdash \Delta_2$ and $\Gamma_3 \vdash \Delta_3$ be the sequents of \lstinline|s1|, \lstinline|s2| and \lstinline|s3|. Now, to check the correctness of \lstinline|s2|, we must verify:
$$\Gamma_1 == \Gamma_2$$
$$\lbrace \text{\lstinline|(f(X) = Y)|}[ \text{\lstinline|Y|}:=\text{\lstinline|f(X)|} ]\rbrace \cup \Delta_2 ==
\lbrace \text{\lstinline|?[Y]: (f(X) = Y)|}\rbrace \cup\Delta_1$$
Where $\phi[X:= t]$ denotes the (capture-avoiding) substitution of $X$ by $t$ in $\phi$. Note that the equality has to be performed modulo alpha-equivalence. This can be done naively in time $\mathcal O (n^2)$, but also efficiently either by using some hash function that is congruent with respect to alpha-equivalence, such as in \cite{maziarzHashingModuloAlphaEquivalence2021}, or more simply by computing a locally nameless normal form for formulas.

For step s3, the same checks need to be done, but additionally, it must be verified that the quantified variable $X$ is not free in the resulting sequent.

The \lstinline|Cut| step is slightly different: because its main formula does not appear in the conclusion, the index indicates instead the position cut formula in the right-hand side of the first premise (this is arbitrary; we could have pointed instead to the cut formula in the left-hand side of the second premise).

Finally, the \lstinline|leftSubst| and \lstinline|rightSubst| rules are a bit more complex. Consider for example the following derivation:
\begin{lstlisting}
fof(a1, axiom, [P(f(a))] --> [])
        
fof(s1,plain, [P(g(b)), (f(a) = g(b))] --> [], 
        inference(leftSubst, [status(thm), 1, $fof(P(Z)), $fot(Z)), [s1]))
\end{lstlisting}
The parameter $1$ points to the equality \lstinline|f(a) = g(b)|. The second and third parameters explain how the substitution is carried, so for general sequents $\Delta_1$ and $\Delta_2$ as above, the check is:

\begin{center}
    $\lbrace$\lstinline|P(Z)|$[$\lstinline|Z|$:=$\lstinline|g(b)|$]$, \lstinline|f(a) = g(b)|$\rbrace \cup \Delta_1 == 
    \lbrace$\lstinline|P(Z)|$[$\lstinline|Z|$:=$\lstinline|g(a)|$]\rbrace \cup \Delta_2$
\end{center}

\subsection{Level 2 Deduction Steps}
\label{sect:lvl2}

Unlike in level 1, level 2 steps are not entirely fixed and are expected to expand over time. They contain more advanced proof steps, which may be more difficult to verify, but for which there should be an available and implemented algorithm eliminating them from a proof. This mechanism allows Level 1 proofs to be rebuildable from a Level 2 proof. We expect that proofs relying on level 2 proof steps will be common in practice: each tool will keep those they accept natively (or can import easily), and eliminate steps they do not support.
At present time, we have implemented three level 2 steps, which were useful to our implementation: left and right simultaneous substitution of equal terms, and congruence closure. These and their parameters are shown in \autoref{tab:SCTPTP_rules_lvl_2}. Another example of a level 2 candidate (not implemented) is an \texttt{NNF} step, which allows deduce a sequent from a premise whose formulas are equivalent, modulo negation normal form.

\renewcommand{\arraystretch}{1.9} 
\begin{table}
    \centering
    
    \begin{tabular}{|c|c|c|l|}
         \hline
        Rule name & Premises & Rule & Parameters \\ \hline  
        \texttt{NNF}  & 1 & \makecell{\AxiomC{$\Gamma \vdash \Delta$} \UnaryInfC{$\Gamma' \vdash \Delta'$} \DisplayProof}  & \makecell[l]{No parameters\\The premise and conclusion are equal \\up to negation normal form} \\ \hline
        \texttt{congruence}  & 0 & \makecell{\AxiomC{} \UnaryInfC{$\Gamma, P(u) \vdash P(t), \Delta$} \DisplayProof} & \makecell[l]{No parameter\\$\Gamma$ contains a set of equalities such that\\$t$ and $u$ are congruent\\
        (actually more general, see bellow)} \\ \hline 
        \texttt{rightSubstMulti}  & 1 & \makecell{\AxiomC{$\Gamma \vdash P(t_1,...,t_n), \Delta$} \UnaryInfC{$\Gamma \vdash P(u_1,...,u_n), \Delta$} \DisplayProof} &  \makecell[l]{\texttt{[i${}_1$, ..., i${}_n$:Int]}: Index of $t_j = u_j$\\on the left   \\ \texttt{P(Z${}_1$, ..., Z${}_n$):Term}: Shape of the\\ formula on the right \\ \texttt{[Z${}_1$, ..., Z${}_n$:Var]}: variables indicating\\where to substitute} \\ \hline   
        \texttt{leftSubstMulti}  & 1 & \makecell{\AxiomC{$\Gamma, P(t_1,...,t_n) \vdash \Delta$} \UnaryInfC{$\Gamma, P(u_1,...,u_n) \vdash \Delta$} \DisplayProof} &  \makecell[l]{\texttt{[i${}_1$, ..., i${}_n$:Int]}: Index of $t_j = u_j$\\on the left   \\ \texttt{P(Z${}_1$, ..., Z${}_n$):Term}: Shape of the\\ formula on the left \\ \texttt{[Z${}_1$, ..., Z${}_n$:Var]}: variables indicating\\where to substitute} \\ \hline   
    
    \end{tabular}
    \caption{Level 2 rules of \sctptp, for one-sided and two-sided sequent calculus.}
    \label{tab:SCTPTP_rules_lvl_2}
\end{table}
\renewcommand{\arraystretch}{1}

Simultaneous substitutions are fairly simple. A simultaneous substitution of $n$ formulas can always be unfolded into $n$ application of \texttt{leftSubst} or \texttt{rightSubst}. Congruence closure is more technical. A \texttt{congruence} step for a sequent $\Gamma \vdash \Delta$ is correct if one of the following cases hold, given all the formulas of the form $s = t$ in $\Gamma$:
\begin{enumerate}
    \item There are two formulas $P(a_1,...,a_n) \in \Gamma$ and $P(b_1,...,b_n) \in \Delta$ such that for all $i$, $a_i$ and $b_i$ are congruent (\texttt{hyp} case).
    \item There are two formulas $P(a_1,...,a_n) \in \Gamma$ and $!P(b_1,...,b_n) \in \Gamma$ such that for all $i$, $a_i$ and $b_i$ are congruent (\texttt{leftHyp} case).
    \item There is a formula $a = b \in \Delta$ such that $a$ and $b$ are congruent (\texttt{rightRefl} case).
\end{enumerate}
This was immediately useful to us because \goeland{} uses \textit{Rigid E-Unification} \cite{gallierTheoremProvingUsing1987, degtyarev1998you} to close branches, which becomes a congruence-like step at proof reconstruction. We implemented a method to eliminate such \texttt{congruence} steps in \sctptp{} proofs using e-graphs, as explained in \autoref{sect:egraph}.


\section{A Library of Utilities for \sctptp{}}
\label{sect:utilities}

To support the \sctptp{} format, we started the development of a library of tools and utilities to parse, print, verify, and transform \sctptp{} proofs. We chose Scala to implement it because it is a high-level language adapted to the task, it is the language of \lisa{} and Princess (which we plan to support in the future), and there already exists a complete \tptp{} parser in Scala, thanks to Alexander Steen\footnote{\url{https://github.com/leoprover/scala-tptp-parser}}. This library of tools is available at \url{https://github.com/SC-TPTP/sc-tptp}.

The library contains the syntactic definition of first-order logic, sequents, and deduction steps from \autoref{tab:SCTPTP_rules_lvl_1_1} to \autoref{tab:SCTPTP_rules_lvl_2}. It also contains a parser (based on the aforementioned one) for \sctptp{} files, as well as a printer and a proof checker, which report incorrect steps. In addition, the library also contains a printer exporting \sctptp{} proofs to \coq{} proofs. Finally, it contains a tool to eliminate level 2 steps (in particular \texttt{congruence}).

\subsection{Proof Export to \coq}
\label{sect:coq}

The translation of \sctptp{} proofs to \coq{} proofs is relatively straightforward, as each sequent calculus step can be translated into a \coq{} lemma. Our translation relies on a one-to-one mapping between \sctptp{} rules and \coq{} lemmas, as exemplified below:

\begin{example} Translation of \texttt{rightAnd} in \coq.
    \begin{lstlisting}[language=coq]
Lemma rightAnd : forall P Q : Prop, (P) -> (Q) -> (P /\ Q).
Proof. intros P Q H. split. auto. auto. Qed.
    \end{lstlisting}
\end{example}

Translation of rules that generate one conclusion is pretty direct. However, as \coq{} is based on \emph{intuitionistic} logic, it only allows the conclusion to have at most one element. Consequently, rules that create two formulas on the right of the sequent must be rearranged to work with the hypothesis, as in the following example: 

\begin{example} Translation of \texttt{rightOr} in \coq.
    \begin{lstlisting}[language=coq]
Lemma rightOr : forall P Q : Prop, ~(~P /\ ~Q) -> (P \/ Q).
Proof. intros P Q H. apply NNPP. intro H1. apply H. split. auto. auto. Qed.
    \end{lstlisting}
\end{example}

This lemma negates the formula, which subsequently allows us to introduce it and generate multiple formulas within the hypothesis. In order to make use of those rules, we need to use \texttt{apply} on the corresponding hypothesis before applying \emph{right} rules, and to end with \texttt{intro}. For instance, the sequent system \emph{right or} rule would generate \texttt{P, Q} in the conclusion of the sequent. Our \coq{} rule enables that by keeping $\mathtt{\neg P}$ and $\mathtt{\neg Q}$ as hypotheses. As such, when needing \texttt{P} or \texttt{Q}, it suffices to apply $\mathtt{\neg P}$ or $\mathtt{\neg Q}$ and then proceed normally.

Moreover, \emph{left} rules require an additional mechanism to be translated into \coq. Indeed, in \coq, two things can be achieved thanks to a formula \texttt{A -> B} in the hypothesis: either we have \texttt{B} in the conclusion and we can generate \texttt{A} in the conclusion, or we have \texttt{A} as a hypothesis and we can generate \texttt{B} in the hypothesis. However, as our proof is built on an \emph{abductive} way, we want to obtain \texttt{A} in hypothesis from a hypothesis \texttt{B}. In order to do that, complementarily to the rule itself, we need to define additional lemmas that ``invert'' the terms in the proof. We thus need to consider a proof with ``holes'', and wait for \coq{} to fill them. 

Let us illustrate it with the \emph{left implies} rule. The sequent rule states that from two premises (one with $\mathtt{\neg P}$ and the other one with \texttt{Q} as hypothesis), we can infer \texttt{P -> Q} as a hypothesis. However, in our proof, we have \texttt{P -> Q}, and so we are not able to deduce anything. Luckily, we can define a lemma that inverts the rule in order to make it applicable with the conclusion $\mathtt{P -> Q}$, leaving a hole in the hypothesis, which we will have to prove later by providing a witness for $\mathtt{\neg P}$ and $\mathtt{Q}$. Those inversion steps are suffixed with \texttt{\_s}. 

\begin{example}  Translation of \texttt{leftImp} in \coq.
    \begin{lstlisting}[language=coq]
Lemma leftImply_s : forall P Q : Prop,
  (~P -> False) -> (Q -> False) -> ((P -> Q) -> False).
Proof. tauto. Qed.

Definition leftImply := fun P Q c hp hq => leftImply_s P Q hp hq c.
    \end{lstlisting}
\end{example}

Finally, the context of the proof (i.e., constant, predicates, functions, etc, converted into \texttt{Parameter}s) is retrieved and added at the beginning of the proof. 
\begin{example} Context for the \emph{Drinker's paradox} proof.
    \begin{lstlisting}[language=coq]
Parameter sctptp_U : Set. (* universe *)
Parameter sctptp_I :  sctptp_U. (* an individual in the universe. *)
Parameter d: sctptp_U -> Prop.
Parameter X_0: sctptp_U.
    \end{lstlisting}
\end{example}

All the lemmas used for the translation can be found in \texttt{SC-TPTP.v} (\url{https://github.com/SC-TPTP/sc-tptp/tree/main/src}). Then, the translation of a full proof can be done by mapping each proof step to its corresponding lemma, using the parameters of the rule.

\begin{example} \coq{} proof of the \emph{Drinker's paradox}~\cite{smullyan1978name}.
    \begin{lstlisting}[language=coq]
(* Add SCTPTP.p *)
Parameter sctptp_U : Set. (* universe *)
Parameter sctptp_I :  sctptp_U. (* an individual in the universe. *)
Parameter d: sctptp_U -> Prop.
Parameter X_0: sctptp_U.

Theorem drinker:  ~(~(exists (X: sctptp_U), (d(X) -> (forall (Y: sctptp_U), d(Y))))).
Proof.
intro H0.
(* [f3] *) apply H0. exists X_0. apply NNPP. intros H1. 
(* [f4] *) apply (leftNotImp  _  _  H1). intros H2 H3.
(* [f5] *) apply H3. intros Sko_0_15. apply NNPP. intros  H4. 
(* [f6] *) apply H0. exists Sko_0_15. apply NNPP. intros H5. 
(* [f7] *) apply (leftNotImp  _  _  H5). intros H6 H7.
(* [f8] *) auto.
Qed.
    \end{lstlisting}
\end{example}

\subsection{Unfolding Congruence Steps}
\label{sect:egraph}

As motivated in \autoref{sect:lvl2}, to have \texttt{congruence} as a level 2 step, we implemented an algorithm unfolding congruence steps into simpler substitution steps. Our approach is based on computing the congruence closure of all subterms of all atomic and negated atomic formulas in the sequent under the equality given left of a sequent. The congruence closure is computed using an e-graph, a dedicated data structure used in automated theorem provers and program optimization. Our implementation of egraph is in particular inspired by \cite{willseyEggFastExtensible2021} and \cite{nieuwenhuisProofProducingCongruenceClosure2005}. 

An e-graph is built on top of a Union-Find data structure, which maintains an equivalence class of terms under a given set of equality (which in an e-graph are either the input equalities or equalities that follow from congruence).
To produce \sctptp{} proofs, our Union-Find data structure is equipped with an explain method, as in \cite{nieuwenhuisProofProducingCongruenceClosure2005}, which when prompted to explain $a = c$ outputs a path $(a, b_1), (b_1, b_2),..., (b_n, c)$ of equalities. We also record whether an edge comes from an external equality or is a congruence. Congruence edges are then recursively justified by the explain method.
\begin{example}
    Consider the following sequent, justified by a \texttt{congruence} step:
    $$
    (a=b, b=c, P(f(a))) \vdash \neg P(f(c))
    $$
The explanation of  $f(a) = f(c)$ is simply $\textit{congruence}(f(a), f(c))$, and recursively the explanation of $a = c$ is $\textit{external}(a, b), \textit{external}(b, c)$. For any two congruent terms, the proof of equality is produced with a constant number of steps for each edge in the path between the two terms.
\end{example}

\section{Interoperability \& Related Tools}
\label{sect:interoperability_experiments}

This section introduces various tools able to deal with the \sctptp{} format. This format is currently used by the \lisa{} proof assistant and the \goeland{} automated theorem prover, as an export format as well as a means of communication between the two tools. A big picture of the \sctptp{} format use case is available in \autoref{fig:pipeline}. 

\begin{figure}
    \centering
    \begin{tikzpicture}
    \node[draw,rectangle, rounded corners=0.1cm] (sctptp) at (0,0) {\sctptp};
    \node[draw,rectangle, below=1cm of sctptp] (sctptp-utils) {\sctptp{} Utils};
    \node[draw,rectangle, left=1cm of sctptp-utils] (goeland) {\goeland};
    \node[draw,rectangle, right=1cm of sctptp-utils] (lisa) {\lisa};
    \node[draw,rectangle, below=1cm of sctptp-utils] (coq) {\coq{} Proof Assistant};
      \draw (lisa) to (2.55,0.5) to (-2.90, 0.5) edge[->, >=stealth, shorten >=1pt] (goeland);
     \draw (goeland) edge[->, >=stealth, shorten >=1pt] (sctptp);
     \draw (sctptp) edge[->, >=stealth, shorten >=1pt] (lisa);
     \draw (sctptp.250) edge[->, >=stealth, shorten >=1pt] (sctptp-utils.110);
     \draw (sctptp-utils.70) edge[->, >=stealth, shorten >=1pt] (sctptp.290);
     \draw (sctptp-utils) edge[->, >=stealth, shorten >=1pt] (coq);
    \end{tikzpicture}
    \caption{Use cases of \sctptp}
    \label{fig:pipeline}
\end{figure}

\subsection{\goeland}

\goeland~\cite{caillerGoelandConcurrentTableauBased2022, cailler2023designing} is an automated theorem prover for first-order logic with equality. It relies on a concurrent proof-search procedure based on the method of free-variable analytics tableaux that allows it to perform a fair branch exploration. The prover is also able to deal with axiomatisable theories thanks to a module of deduction modulo theory \cite{dowek2003theorem}, to deal with polymorphic types, and to produce machine-checkable proofs in \coq, \lambdapi{} and \lisa. 

\subsection{\lisa}
\lisa~\cite{guilloudLISAModernProof2023} is a proof assistant based on first-order logic, with set-theoretic foundations. Its proof system is inspired by sequent calculus, with additional built-in proof steps to allow for a more efficient representation of typical transformation, such as the substitution of equivalent formulas, simultaneous substitution, quantified substitution, and transformation modulo orthologic \cite{guilloudFormulaNormalizationsVerification2023, guilloudOrthologicAxioms2023}. All the proof steps from \autoref{tab:SCTPTP_rules_lvl_1_1} to \autoref{tab:SCTPTP_rules_lvl_1_5} are readily convertible to \lisa{}'s Kernel steps.

We implemented a printer, that exports queries about a conjectured sequent as a \sctptp{} problem file (but really a \tptp{} file, since it contains no proof), and a parser for \sctptp{} proofs directly to \lisa{}'s Kernel proofs. This allowed us to implement a proof tactic in the user interface, directly using \goeland{} to justify steps prompted by a \lisa{} user, as in the next example.
\begin{example} The \goeland{} tactic in \lisa{}, proving the drinkers problem.
    \begin{lstlisting}[language=lisa]
val drinkers = Theorem(!*$\exists$*!(x, !*$\forall$*!(y, Q(x) ==> Q(y)))) {
    have(thesis) by Goeland
}
    \end{lstlisting}
    
\end{example}

\section{Conclusion}

Our goal with \sctptp{} is to provide a common format to allow sequent-based tools to communicate and exchange proofs, but the adoption of such a format entirely relies on the involvement of the community. In order to increase its potential user base, we plan to expand this format in multiple directions.

The first one will be to increase the number of tools able to deal with this format, starting with tableau-based theorem provers such as \princess~\cite{princess08} or \zenon~\cite{bonichonZenonExtensibleAutomated2007, delahayeZenonModuloWhen2013}. We also want to support the Connection Calculus~\cite{ottenConnectionCalculiAutomated2013, kaliszykEfficientLowLevelConnection2015}, related to Tableaux and used by the Connect++ ATP~\cite{holdenConnectNewAutomated2023} and unify with existing work to export proofs from connection calculi to \tptp. Longer term, we are interested in generalizing our tool to resolution, but this requires additional steps in order to make resolution proofs readily machine-checkable. 

Our proof system can be expanded in many ways, which we hope to explore in the future. An important extension would be the support for typed first order logic. In order to do this, our format needs to be extended to fit with TFF~\cite{blanchette2013tff1}. 
The addition of theory reasoning is also of interest. Theory rules should be possible to add as high-level proof steps, that could be either exported as-is in a proof assistant (if this later is able to deal with the theory), or unfolded in the same way as equality reasoning. Deskolemization~\cite{hermant2013syntactic, rosain2024desko} is also a strong candidate step.
We plan to extend our proof-producing module to export proofs to other proof assistants, such as \lambdapi~\cite{assaf2016dedukti}, \isabelle~\cite{nipkow2002isabelle} or \lean~\cite{avigadTheoremProvingLean}. 

While developping SC-TPTP, we have also encountered limitations within the \tptp{} syntax. As part of our commitment to continuous improvement, we are eager to offer suggestions for refinement. These suggestions may include introducing an \textit{exists unique} quantifier, permitting $n$-ary conjunctions and disjunctions, or proposing a specialized character for expressing identifiers in \tptp{} files, allowing to bind and reuse formulas (and other syntactic expressions). By addressing these limitations, we aim to fortify the foundation of our framework for the collective benefit of the community.

\paragraph{Acknoledgment} This publication is based upon work from COST Action EuroProofNet, supported by COST (European Cooperation in Science and Technology, \url{www.cost.eu})

\bibliography{sguilloudZotero}













\end{document}